\documentclass[12pt]{article} 

\usepackage{hyperref}
\usepackage{url}
\usepackage{setspace}
\usepackage{amsmath,amssymb,amscd,braket}
\usepackage{amsfonts}
\usepackage{color}
\usepackage{graphicx}
\usepackage{cite}
\usepackage{bbm}

\numberwithin{equation}{section}

 \newcommand{\bea}{\begin{eqnarray}}
\newcommand{\eea}{\end{eqnarray}}
\newcommand{\be}{\begin{equation}}
\newcommand{\ee}{\end{equation}}
\newcommand{\ba}{\begin{align}}
\newcommand{\ea}{\end{align}}

\newcommand{\tr}{ {\rm Tr}}
\newcommand\rref[1]{(\ref{#1})}

\newlength{\slength}
\renewcommand{\title}[1]{\vbox{\center\LARGE{#1}}\vspace{5mm}}
\renewcommand{\author}[1]{\vbox{\center#1}\vspace{5mm}}
\newcommand{\address}[1]{\vbox{\center\footnotesize\em#1}}
\newcommand{\email}[1]{\vbox{\center\footnotesize\tt#1}\vspace{5mm}}

\begin{document}

\begin{titlepage}

\begin{center}

\hfill \\
\hfill \\
\vskip 1cm

\title{Baby Universes in AdS$_3$}

\author{Alexandre Belin$^{a,b}$ and Jan de Boer$^c$
}

\address{
${}^a$Dipartimento di Fisica, Universit\`a di Milano - Bicocca \\
I-20126 Milano, Italy

\vspace{1em}
${}^b$INFN, sezione di Milano-Bicocca, I-20126 Milano, Italy

\vspace{1em}
$^c$ Institute for Theoretical Physics \\
PO Box 94485, 1090GL Amsterdam, Netherlands
}

\email{alexandre.belin@unimib.it, J.deBoer@uva.nl}

\end{center}

\abstract{

We discuss Euclidean geometries in AdS$_3$ whose Lorentzian slicing gives rise to closed baby universes with a spatial geometry given by genus $g\geq 2$ surfaces. Our setup only involves a two-dimensional holographic CFT defined on a higher genus Riemann surface and thus provides a well-posed alternative to shell states whose microscopic duals are less well understood. We find that geometries giving rise to baby universes are always subdominant. It follows that the baby universe does not provide a semi-classical description of the state since it is encoded in an exponentially suppressed part of the wave function. We then apply a prescription developed in \cite{Belin:2025wju} to make the baby universe geometry the leading saddle. In the process, the CFT state becomes mixed, in agreement with the qualitative gravitational picture. We show that the fluctuations in the baby universe are small, even at fixed central charge, making the geometry reliable in the semi-classical limit. Finally, we discuss the interpretation of this mixed state in pure gravity from the perspective of the Virasoro TQFT.

}

\vfill

\end{titlepage}

\eject

\tableofcontents

\section{Introduction}

The gravitational path integral is a powerful tool to study the inner workings of quantum gravity, constantly providing new insights while simultaneously raising new puzzles. In recent years, a refined understanding of spacetimes with more complicated topologies - in particular  wormholes - has led to considerable progress on the black hole information paradox \cite{Penington:2019kki,Almheiri:2019qdq,Saad:2018bqo,Saad:2019lba}. At the same time, the inclusion of such topologies in the gravitational path integral leads to puzzles, most notoriously the factorization puzzle \cite{Maldacena:2004rf}. 

The current understanding of the appearance of such geometries is that they are the emergent holographic manifestation of coarse-grained statistical correlations between black hole microstates \cite{Pollack:2020gfa,Belin:2020hea,DiUbaldo:2023qli,Haehl:2023tkr,Boruch:2025ilr}, where the coarse-graining is meant to follow from a maximum ignorance principle \cite{deBoer:2023vsm}. This viewpoint applies to standard AdS/CFT pairs where no averaging over theories is involved. But for very simple gravitational theories, the boundary dual may be an actual average over theories. This has been established for JT gravity \cite{Saad:2019lba,Jafferis:2022wez} and accumulating evidence is pointing for a similar situation in the case of pure quantum gravity in AdS$_3$ \cite{Cotler:2020ugk,Chandra:2022bqq,Collier:2023fwi,Belin:2023efa,Collier:2024mgv,Jafferis:2025vyp,deBoer:2025rct,Chandra:2025fef,Hartman:2025ula,Belin:2025qjm}.

In this paper, we will address one of the puzzles that has resulted from the consideration of wormholes: baby universes. Slicing Euclidean geometries of complicated topologies to obtain states living on a time slice reveals very rich physics. In many cases, this leads to several disjoint universes, some of which may be closed. In a fixed theory and on general grounds, the gravitational path integral has provided evidence that the Hilbert space of closed universes should be one-dimensional \cite{Marolf:2020xie}.\footnote{This is meant to hold as a non-perturbative statement. Semi-classically, one can engineer large Hilbert spaces by adding features (or observers) with respect to which we can dress local operators, and for which the semi-classical Hilbert space can be large, see for example \cite{Chandrasekaran:2022cip,Bahiru:2022oas,Bahiru:2023zlc,Jensen:2023yxy,Chen:2024rpx,Speranza:2025joj}.} Nevertheless, certain explicit constructions seem to be in tension with this statement.

\begin{figure}[h]
\centering
  \includegraphics[width=.95\linewidth]{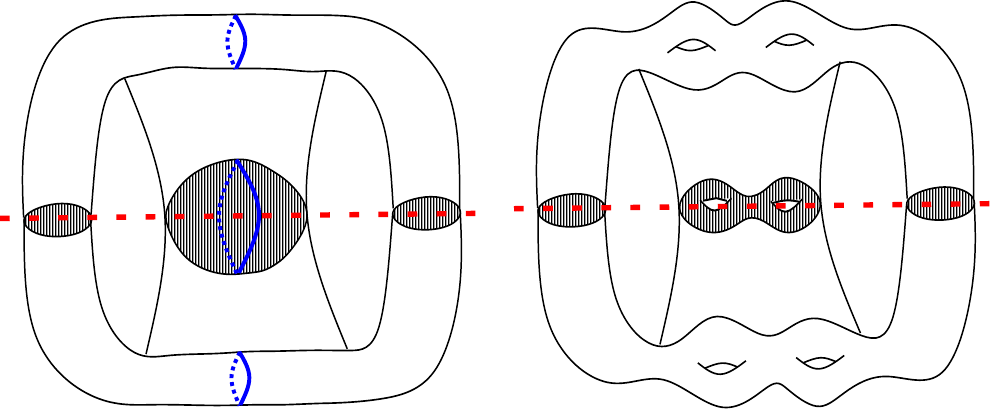}
\caption{The difference between the AS${}^2$ setup (left) and our setup (right). These geometries correspond to the overlap of the state with itself. In the AS${}^2$ setup, the state is prepared by a path integral on a cylinder with a shell inserted. The shell propagates inside the Euclidean geometry, and the $t=0$ slice corresponds to two disconnected AdS regions, entangled with a baby universe on which the shell lives. In our setup, the shell is replaced by topology. Unlike shells, the geometry shown here is \textit{not} the only geometry with these boundary conditions, there are others, and these other saddles dominate over the one that is represented here.}
  \label{fig:comparison}
\end{figure}

One case that has made these puzzles sharp is a construction due to Antonini, Sasieta and Swingle (AS${}^2$). It comes from considering a bulk theory given by general relativity coupled to pressureless dust. This matter propagates on thin-shells, and the gravitational configurations involving these shells are called shell geometries, and the CFT duals shell states. These geometries have been used to study black hole collapse \cite{Anous:2016kss} (albeit with different kinematics) as well as the over-completeness of semi-classical states and the associated finiteness of the black hole Hilbert space \cite{Balasubramanian:2022gmo,Balasubramanian:2022lnw,Climent:2024trz,Balasubramanian:2025gtv,deBoer:2025tmh,Balasubramanian:2025zey,Balasubramanian:2025hns}. The setup of AS${}^2$ is exactly the same as that of the one to study over-completeness, but in a different regime. At high energies, the shell states give rise to black hole geometries with long throats behind the horizon. AS${}^2$ performed a Euclidean cooling of these states, such that they go through a type of Hawking-Page phase transition. Below the Hawking-Page phase transition, the entropy and energies scale as $N^0$, but can be large. The bulk geometry consists of two disconnected AdS spacetimes, entangled with a baby-universe on which the shell lives, see Fig. \ref{fig:comparison}.

Antonini and Rath (AR) showed that this setup raises a puzzle \cite{Antonini:2024mci}. The CFT state preparing the shell appears to be a pure state in the CFT. Because the energy and entropy of the state scales as $N^0$, the microstates making up this entropy are well-understood: they are the states of the Fock space of perturbative fields propagating on AdS. Whatever the shell state is, it must be expressible as a pure state living in the Hilbert space spanned by these low-energy Fock space states. The bulk geometry with the baby universe suggests otherwise: the state of two AdS spaces is mixed, as it is entangled with the baby universe. This is the puzzle, which has triggered much debate on the true dimensionality of the baby universe Hilbert space and on whether this baby universe is semiclassical \cite{Engelhardt:2025vsp,Engelhardt:2025azi,Harlow:2025pvj,Antonini:2025ioh,Gesteau:2025obm,Liu:2025cml,Kudler-Flam:2025cki,Mori:2025jej}.

In this paper, we will propose a general mechanism by which puzzles of the AR type are avoided. We believe that this mechanism is generic for states made using the CFT path integral, but we will illustrate it for states in AdS$_3$/CFT$_2$ obtained by performing the CFT path integral on higher genus Riemann surfaces. For such states, we establish the following:
\begin{enumerate}
\item Geometries giving rise to baby universes exist, and are in fact generic.
\item These geometries are never the leading saddles contributing to the overlap of the associated path integral states. The leading saddle in the low-energy AS${}^2$-like limit is a geometry without any baby universe, which is a partially entangled state, but is pure on two copies of the CFT. The saddles with baby universes are exponentially suppressed and as such, while the baby universe exists in some exponentially small part of the wave function, it cannot be thought of as a semi-classical description of the states prepared by the CFT path integral.
\item It is possible to perform a microscopic operation on the state, in a logic similar to that introduced in \cite{Belin:2025wju}, to engineer a new state whose leading order geometry contains a baby universe.\footnote{By leading order, here we mean that the trace of the (unnormalized) density matrix gives to leading order the gravity partition function on the baby universe geometry.} In the CFT, this new state is built by a particular microscopic prescription, and no longer corresponds to a state prepared by a Euclidean path integral. The state is explicitly mixed, in accordance with the presence of the baby universe in the bulk. Thus, there is no AR puzzle in our setup.
\item Within this modified state, the wave-function coefficients have small fluctuations, unlike those of shell states \cite{Kudler-Flam:2025cki}. This follows from the fact that our state is not prepared by a CFT Euclidean path integral. Given that fluctuations are small, we can view the baby universe as truly semi-classical. Our prescription is valid at large but fixed $N$, and does not require any averaging over $N$, as advocated for in \cite{Liu:2025cml,Kudler-Flam:2025cki}, but it would be interesting to understand the connection better.
\end{enumerate}

While we believe that our findings are generic properties of states prepared by the CFT path integral, they do not directly resolve the tension for shell states as presented by AR. For such states, we believe that the tension resides in a proper microscopic definition of the CFT states.\footnote{See \cite{SYKtoappear} for a proper microscopic treatment of shell-like states in the SYK model .} The setup for shell states is not a standard application of the GKPW dictionary with bulk matter, but rather corresponds to the insertion of many (order $N$) light operators whose collective effect is assembled into an effective description in terms of pressureless dust. We believe that a proper CFT investigation of the contributions to the many point correlation function of all these light operators would reveal other contributions beyond the one captured by the bulk shell, but it is not clear whether they can all be thought of as being cleanly geometric.\footnote{See \cite{Chandra:2024vhm} for a bootstrap/statistical treatment of these shells in 3D gravity.} From this perspective, our setup has the advantage of being a standard application of the GKPW dictionary: the CFT state simply comes from turning on a source for the stress-tensor, by putting the CFT on a more complicated topology.

The interpretation of the mixed state whose leading bulk dual is the baby universe is as follows. In pure 3D gravity\footnote{Most of the results that we establish are based on calculations in pure 3D gravity, but we discuss adding matter in the discussion.}, the baby universe Hilbert space is not a standard local QFT Hilbert space on a closed spatial section. This follows from the fact that the bulk theory is topological. Nevertheless, there still is a Hilbert space for the baby universe, which is an auxiliary Hilbert space corresponding to the space of conformal blocks, familiar from TQFTs. This auxiliary Hilbert space is labeled by Virasoro representations and provides a purification of the mixed state on the two AdS universes. We also discuss the canonical purification that follows from the GNS construction, which can be prepared by performing the bulk path integral on a complicated topology with four AdS universes. The auxiliary Hilbert space can be viewed as an entanglement bottleneck between two pairs of AdS universes, the original two and the two purifying the state. We speculate on a link between this auxiliary Hilbert space and observers.

The rest of the paper is organized as follows: in Section \ref{sec:2}, we describe our construction in AdS$_3$. We show that geometries with baby universes exist, but that they are never the dominant saddle. In Section \ref{sec:3}, we perform the microscopic operation on the state that makes such geometries dominant. We show that the CFT state becomes mixed, in agreement with the presence of the baby universe. We also discuss the interpretation of the mixed state in pure 3D gravity. We conclude with some open questions in Section \ref{sec:4}.

\section{The setup in AdS${}_3$ gravity \label{sec:2}}

\begin{figure}[h]
\centering
  \includegraphics[width=.8\linewidth]{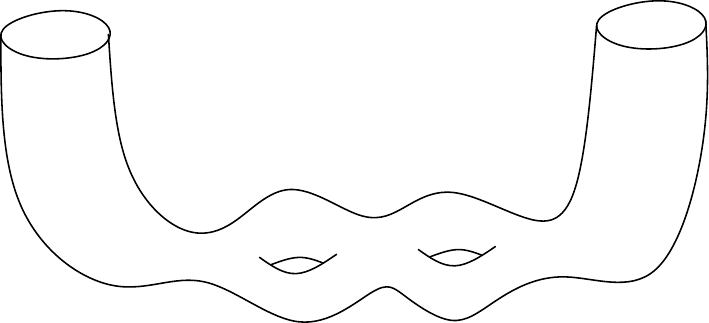}
\caption{The preparation of the state $\ket{\frac{1}{2} [g=5]}$ using the CFT path integral. The path integral is performed over half of a genus-5 surface, and this preparation yields an entangled state in the tensor-product of two copies of the CFT Hilbert space.}
  \label{fig:halfgenus5}
\end{figure}

In this paper, we will consider a setup in AdS$_3$/CFT$_2$ that nicely parallels the discussion of shell states originally presented in \cite{Antonini:2023hdh}. To be concrete, we will be interested in a CFT state that we call $\ket{\frac{1}{2} [g=5]}$.\footnote{This example is chosen for concreteness, and we comment on the many possible generalizations such as higher genera in the discussion.} This state is obtained in the CFT by performing the Euclidean path integral over half of a genus-5 surface, as depicted in Fig. \ref{fig:halfgenus5}. The genus-5 surface has been cut so that the cut intersects two circles. This implies that the CFT state lives in the Hilbert space of the tensor product of two copies of the CFT Hilbert space
\be
\ket{\frac{1}{2} [g=5]} \in \mathcal{H}_L \otimes \mathcal{H}_R \,.
\ee
In general, this state will be entangled.

In holography, the description of this state depends on the moduli of this half genus-5 surface. The bulk physics are governed by the leading saddle of the overlap of the state with itself, which in this case forms a genus-5 partition function 
\be \label{overlap}
\braket{\frac{1}{2} [g=5] \Big| \frac{1}{2} [g=5] } = Z_{g=5} \,.
\ee
For CFT path integral states, the bulk dual of a  state is found by selecting the leading saddle to this partition function, finding a time-reflection symmetric slice and performing an analytic continuation to Lorentzian signature to obtain initial data for the gravitational theory, see \cite{Skenderis:2008dh,Botta-Cantcheff:2015sav,Marolf:2017kvq,Belin:2018fxe} or \cite{Belin:2025wju} for a short overview.

The most common regime studied for these types of states is the one relevant for multi-boundary black holes, see \cite{Balasubramanian:2014hda}. This corresponds to the regime where the two vertical cylinders are very short. In that case, the metric for the bulk geometry dominating \rref{overlap} is
\be \label{highergBH}
ds^2= d\tau^2 + \cosh^2 \tau \ d\Sigma_{\tilde{g}}^2 \,,
\ee
where $d\Sigma_{\tilde{g}}^2$ is the constant negative curvature metric on the surface shown in Fig.  \ref{fig:halfgenus5}, namely the half-genus-5 surface (or equivalently, we can think of it as a genus-2 surface that itself has two asymptotically AdS boundaries attached to it).  
This three-dimensional metric admits a Lorentzian continuation at the slice $\tau=0$. The Lorentzian metric reads
\be \label{BHphase}
ds^2= -dt^2 + \cos^2 t \ d\Sigma_{\tilde{g}}^2 \,,
\ee
and it covers the Wheeler-de Witt patch of the $t=0$ slice. 
On this slice, we find the half genus-5 surface, which is a generalization of an Einstein-Rosen bridge to a geometry with topology behind the left and right horizons. This state is connected in the bulk, so it has entanglement of the order of the central charge, but it is not thermally entangled like the thermofield-double state \cite{Maldacena:2001kr}.

 \begin{figure}[h]
\centering
  \includegraphics[width=.8\linewidth]{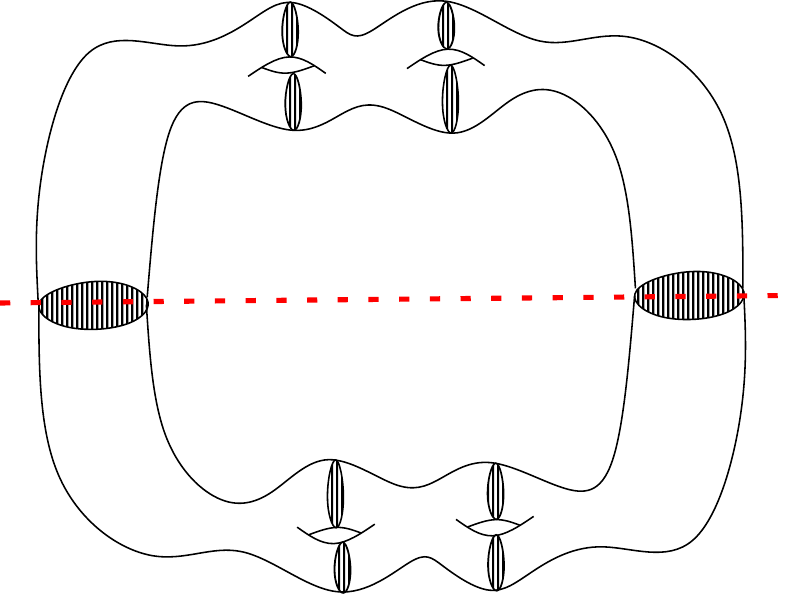}
\caption{The geometry of the handlebody at genus-5. The boundary of this bulk three-dimensional geometry is the genus-5 surface that corresponds to the overlap \rref{overlap}. The geometry is obtained by filling in a choice of cycles, as shown here. On the $t=0$ $\mathbb{Z}_2$-symmetric slice where we cut the overlap to get the state, we simply have two disconnected disks. The quantum fields on these disks are entangled, just like in thermal AdS, but the entanglement pattern is more complicated than the familiar thermal one.}
\label{fig:handlebody}
\end{figure}

In this paper, we will study a different regime of moduli space: we will study the regime where the two cylinders are very long. Making the two tubes longer and longer decreases the expectation value of the energy, and starting from the black hole geometry with topology behind the horizon, elongating the two tubes will lead to a Hawking-Page phase transition where the dominant geometry is now a generalization of thermal AdS. In this regime, the bulk geometry is just the obvious filling of the surface that prepares the state. This is represented in Fig. \ref{fig:handlebody}.  The bulk geometry at $t=0$ is simply two disconnected AdS disks. The state still has non-trivial (but importantly order $c^0$) entanglement, and the pattern of entanglement is more generic than that in thermal AdS which is a thermal entanglement pattern for the light bulk fields.

So far, we have considered two types of geometries contributing to the overlap \rref{overlap}. Both of these geometries are genus-5 handlebodies, meaning that they are fillings of the genus-5 surface. They correspond to different choices of cycles that become contractible in the bulk. In fact, we have only cared about two of the cycles of the surface: those that correspond to the two vertical tubes. We have been agnostic about the fate of the other cycles, and there are also Hawking-Page phase transitions as those moduli are varied. For our purposes, these are not relevant, and we will always pick the leading saddle in terms of those cycles, and focus our attention on the fate of the two vertical cylinders. The first choice leading to the metric \rref{BHphase} is the case where the two cycles of the vertical cylinders are not contractible, and the phase described in the previous paragraph corresponds to the case where they are both contractible.

\begin{figure}[h]
\centering
  \includegraphics[width=.8\linewidth]{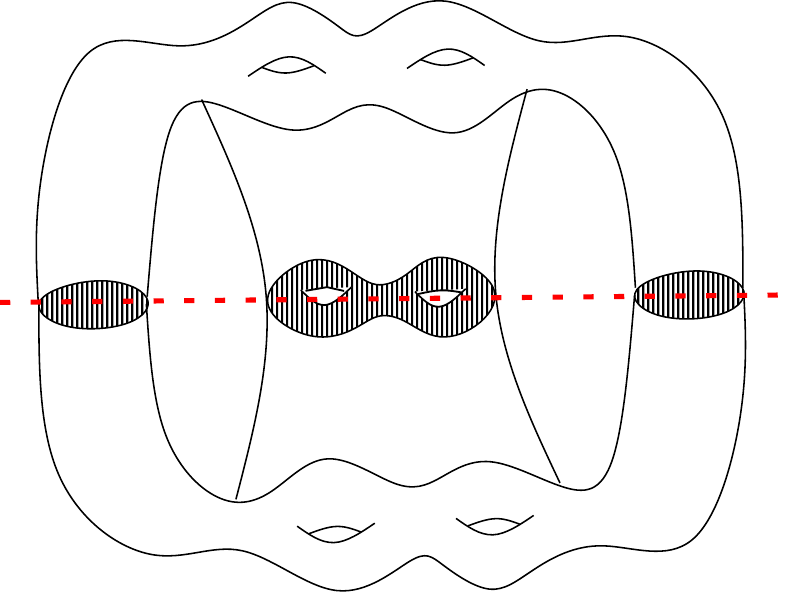}
\caption{The geometry of the non-handlebody at genus-5. The boundary of this bulk three-dimensional geometry is the genus-5 surface that corresponds to the overlap \rref{overlap}. Unlike a handlebody, the cycles of the bottom and top genus-2 surfaces are not contractible. Instead, the bottom and top genus-2 surfaces are connected by a Maoz-Maldacena wormhole. On the $\mathbb{Z}_2$-symmetric slice drawn as the dashed line, we can slice the geometry and continue it to Lorentzian signature. This will lead to two asymptotically AdS disks, and a closed universe made by a genus-2 surface.}
\label{fig:non-handlebody}
\end{figure}

But there are also other geometries which are not of the handlebody type. These would not exist (at least not as on-shell saddles of 3D gravity) if the state was the thermofield double-state, but they do exist for the overlap of our state. One of these geometries will be of particular interest to us. It corresponds to taking the Maldacena-Maoz wormhole \cite{Maldacena:2004rf}, and gluing onto it two solid tubes that connect the two asymptotic boundaries. The metric of this geometry cannot be written down in a simple way, but we show a cartoon of this geometry in Fig. \ref{fig:non-handlebody}. We emphasize that it is an on-shell solution to Einstein's equations, i.e. it is locally AdS and is obtained from Euclidean AdS$_3$ by quotienting with respect to the appropriate discrete subgroup of the isometry group. Alternatively, it can be obtained from the geometry of the handlebody using genus two surgery as in \cite{deBoer:2025rct}. 

In this geometry, the geometry of the $t=0$ slice consists not only of two asymptotically AdS disks, but also contains a closed geometry: the genus-2 surface in the middle. This is precisely the baby universe. The metric for the baby universe is again
\be
ds^2=-dt^2 + \cos^2 t \ d\Sigma_{g}^2 \,,
\ee
but instead of having the metric on a genus-2 surface with two punctures that become asymptotically AdS regions outside of the horizon, here the Riemann surface is a closed genus-2 surface. If one performs (forward or backward) Lorentzian time evolution, the closed universe will collapse, so this geometry describes a big bang/big crunch closed cosmology. There are also two global AdS factors that correspond to the two AdS universes.

\subsection{The CFT state and dominance of saddles}

The state $\ket{\frac{1}{2} [g=5]}$ can expressed purely in CFT terms.\footnote{In this paper, we will work with \textit{unnormalized} states, such that the overlap of a gravitational state with itself gives the exponential of the on-shell action in the bulk. For overlaps of different states, it would be important to take into account the normalization factors.} 
It reads
\be
\ket{\frac{1}{2} [g=5]}=\sum_{i,j,k,l,m,n,o} C_{ikl}C_{kmn}C_{lmo}^*C_{noj}^* \mathbb{O}_{i,j}(\Omega,h_k,h_l,h_m,h_n,h_o) \ket{0}_L \otimes  \ket{0}_R \,.
\ee
The notation is as follows. The $C_{\alpha \beta \gamma}$ are OPE coefficients of the CFT. The sum over the indices $i,j,k,l,m,n,o$ runs over all primary states of the CFT. $\Omega$ represents the period matrix of the Riemann surface, and therefore encodes all the moduli of the surface. The operator $\mathbb{O}_{i,j}(\Omega,h_k,h_l,h_m,h_n,h_o)$ is what we will call a half genus-5 OPE block. It contains both primaries and descendants, but restricting to primaries it is of the form\footnote{In principle, the operator $O_j$ is actually CPT conjugated, but we will omit this for simplicity.}
\be
\mathbb{O}_{i,j}(\Omega,h_k,h_l,h_m,h_n,h_o) \Big|_{\text{primary}} = f(\Omega, h_i,h_j,h_k,h_l,h_m,h_n,h_o)) O_i ^L \otimes O_j^R
\ee
where $f$ is a function that depends on the moduli and weights of all the operators. For internal operators (that is, excluding the primaries $i$ and $j$), all descendants have already been resummed in the function $f$. The full operator $\mathbb{O}_{i,j}$ also includes Virasoro descendants of the primaries $O_i$ and $O_j$. The expectation value of the square of an OPE block is the genus-5 Virasoro block (squared for left and right movers). 
\be
\bra{0}\otimes \bra{0} \mathbb{O}^\dagger_{i,j}(\Omega,h_k,h_l,h_m,h_n,h_o)\mathbb{O}_{i,j}(\Omega,h_k,h_l,h_m,h_n,h_o)\ket{0}\otimes \ket{0} = |\mathcal{F}_{g=5}|^2 \,.
\ee
We have suppressed the choice of channel and the dependence on the weights and moduli in the genus-5 block. 

In 3D gravity, the genus-5 handlebodies correspond to identity Virasoro blocks in various channels. In our case, we will always be interested in taking the length of the vertical cylinders to be sufficiently long such that the vacuum dominates, up to exponentially small corrections. This means that we will be interested in the vacuum OPE block $\mathbb{O}_{\mathbbm{1},\mathbbm{1}}$. In the rest of this paper, we will consider the case of pure 3D gravity, where the only light operator is the identity, and the rest of the operators are above the black hole threshold. If the gravitational theory had matter, there would be other blocks contributing, corresponding to the matter fields. We will come back to this in the discussion. In the long cylinder limit, the state becomes
\be \label{lowTstate}
\ket{\frac{1}{2} [g=5]}\approx \sum_{l,m,n} |C_{lmn}|^2\mathbb{O}_{\mathbbm{1},\mathbbm{1}}(\Omega,h_l,h_m,h_n,) \ket{0}_L \otimes  \ket{0}_R \,.
\ee

Even in this region of moduli space, there is still a lot of freedom for the remaining moduli. They will not play a particularly important role in what follows, but to simplify some expressions, we will take the moduli to be such that the dominant contributions come from heavy operators in the sum over $l,m$ and $n$. At the level of the dominant handlebody in this limit, this is equivalent to picking the identity in a cross channel.\footnote{Note that this is also a different phase than the one drawn in Fig. \ref{fig:handlebody}.} But we will not only be interested in the handlebody, so we will keep the sum over all heavy operators in the original channel instead.

We can now understand where the different geometries come from, thinking about the statistical distribution of OPE coefficients in holographic CFTs \cite{Collier:2019weq,Belin:2020hea,Belin:2021ryy,Anous:2021caj}, which for pure gravity can be made precise \cite{Chandra:2022bqq,Belin:2023efa,Collier:2023fwi,Collier:2024mgv,Hartman:2025cyj}. The OPE coefficients of the CFT have an approximately Gaussian distribution. With this approximation we can see the appearance of the two different geometries we have discussed. The overlap reads
\be
\sum_{i,j,k,l,m,n} |C_{ijk}|^2  |C_{lmn}|^2 |\mathcal{F}_{g=5}|^2 \,.
\ee
The handlebody corresponds to doing a Gaussian contraction between the two OPE coefficients with indices $i,j,k$, and a separate contraction for the indices $l,m,n$, and setting the OPE coefficients equal to the DOZZ formula \cite{Chandra:2022bqq}. The non-handlebody with the baby universe arises from a cross-Wick contraction between each OPE coefficient with indices $i,j,k$ with one with indices $l,m,n$. This is a slight generalization of the Maoz-Maldacena wormhole, where two extra tubes have been added to enhance it into a genus-5 surface. 

In the case of the non-handlebody, the six sums over heavy operators collapse to only three sums, so the action coming from this geometry is exponentially suppressed compared to that of the handlebody. Up to performing the sums with Boltzmann factors coming from the conformal blocks, we expect the non-handlebody to be suppressed schematically by a factor of $e^{-3S(E)}$ where $S(E)$ is the entropy evaluated at the energies at which the sums over heavy operators are peaked.

\subsection{No tension with holography and comparison to AS${}^2$}

We will now explain why our setup, although very closely connected to that of AS${}^2$, does not present any tension with holography. What we found is that the state $\ket{\frac{1}{2} [g=5]}$, for the choices of moduli that we discussed, has as dominant geometry contributing to the overlap a handlebody geometry. The  Lorentzian bulk interpretation of this geometry is that of two disconnected AdS spacetimes, supporting an entangled state of the quantum fields (in pure gravity, this is the state of Virasoro descendants). This geometry does not have any baby universe, and hence there is no puzzle similar to that of AR. On the other hand, there \textit{is} a contribution to the overlap (and hence to the wave-function of the state) which corresponds to two AdS spacetimes entangled with a baby universe, in this case a genus-2 closed universe. But this contribution is exponentially suppressed in the large central charge limit. In this sense, there is no real semi-classical notion in which we can discuss the baby universe \textit{on its own} - at least not in this state, we will discuss a way to achieve this in the next section - and thus there is no semi-classical baby universe. 

One may wonder whether one should be upset about the existence of the baby universe at all, even if it is exponentially suppressed. This implies that the state on the two AdS copies is not exactly pure, but rather has an entanglement of order $e^{-c}$. This is quite similar to the case of the Page curve: the Hawking saddle still exists, so the entropy of the radiation after evaporation is not exactly pure, but rather has exponentially small corrections \cite{Almheiri:2020cfm}. In our setup, we should not be bothered by this. Exponentially small fluctuations (which can be thought of as pseudo-randomness in the light data coming from a coarse-grained description of the high energy part of the theory) is certainly acceptable, and even expected from the bootstrap. A simple example of this is the \textit{exact} anomalous dimension of a double-trace non-BPS operator. Such an anomalous dimension will have a $1/c$ expansion and exponentially small corrections, which are tied to heavy black hole operators, and the magnitude/sign of these corrections can have erratic fluctuation. So on general grounds, we do expect exponentially small random signals, even in the low-energy data.

\subsection*{Comparison to AS${}^2$ and the AR puzzle}

We can now compare our set up to that of AS${}^2$ to see the differences. If we think about the wave-function coefficients of our state, their general structure is as follows. For simplicity, we will discuss only the coefficient that determines the overlap with the vacuum. We therefore consider
\be
A_{\bra{0} \otimes \bra{0}} \equiv \bra{0} \otimes \bra{0} \ket{\frac{1}{2} [g=5]} \,.
\ee
We can write a general ansatz
\be
A_{\bra{0} \otimes \bra{0}}\approx a_0 + e^{-c} a_1 \,.
\ee
Here, the coefficient $a_0$ is computed semiclassically from the handlebody geometry (where there is no baby universe), and there is no randomness or fluctuation in this quantity. The coefficient $a_1$ represents a quantity which is not directly computable from the overlap $A_{\bra{0} \otimes \bra{0}}$, but whose norm can be computed semiclassically from the square of the overlap. Since only the norm is semi-classically accessible, $a_1$ is thought to have a random phase, and we can write
\be
a_1\Big|_{\text{semi-classical}}=0
\ee
The semi-classical computation of the norm of $a_1$ involves a two-boundary computation: $A_{\bra{0} \otimes \bra{0}}^2$. Alternatively, it is also extractable from the overlap of the state $ \ket{\frac{1}{2} [g=5]}$ with itself. So the general structure of the wave-function coefficients of our states is that there is a non-fluctuating piece coming from the leading saddle, and an exponentially small erratic piece coming from the geometry with a baby-universe.\footnote{One way to achieve a situation similar to the shell setup but with topology, would be to explicitly by hand add random phases $e^{i \phi_{l,m,n}}$ in the state \eqref{lowTstate}. Due to the random phases, we would now have $a_0=0$, and the fluctuating piece becomes dominant. But this is not a CFT path integral prescription, it is microscopic fine tuning of the wave function coefficients. Moreover, we expect the fluctuations in such states to be large.} Notice that subleading saddles contributing to the single overlap $A_{\bra{0} \otimes \bra{0}}$ could also produce exponentially small corrections, but since these do not fluctuate they should be thought of as contributing to $a_0$ and not to the erratic quantity $a_1$.

In the shell setup of AS${}^2$, the bulk shell is built in such a way that it leads to
\be
a_0^{\text{shell}}= 0 \,.
\ee
This follows from the fact that a shell of pressureless dust cannot contract on itself: there will never be a saddle with a single shell insertion on the boundary. Because of this, the erratic piece becomes the leading effect. But as discussed in the introduction, the CFT state dual to the shell geometry is not very well understood, and it is not clear how to obtain it from the standard CFT path integral. It may involve explicit coarse-graining/averaging over heavy CFT operators, and as such is not a standard realization of the GKPW dictionary of AdS/CFT. In our state, which is a standard Euclidean path integral state, we find a leading non-erratic piece, which is why there is no obvious mismatch or puzzle with the standard AdS/CFT dictionary.

We believe that the structure found in our state is generic when preparing states with the Euclidean path integral: there will always be self-contracting solutions, and these will dominate in the long-tube/low-temperature limit. It is an interesting (but most likely difficult) problem to attempt to prove such a statement.

\section{Making the baby-universe dominant \label{sec:3}}

A natural question to ask is whether there is any way to make the geometry with a baby universe the leading geometry contributing to the overlap of some state. We will see that the answer is yes, albeit at a price. To this end, we will follow a prescription introduced in \cite{Belin:2025wju}. There, the interest was in a different slicing of the Maoz-Maldacena wormhole: one that in Fig. \ref{fig:non-handlebody} would correspond to a vertical slice through the wormhole. In that context, it was possible to find a pure CFT state $\ket{\psi_{\text{WH}}}$ such that the leading contribution to $\braket{\psi_{\text{WH}}|\psi_{\text{WH}}}$ is the Maoz-Maldacena wormhole.

We will follow a similar logic to find a state whose leading semi-classical description will correspond to the geometry with a baby universe. The prescription will be slightly more subtle and, as we will see, will lead to a mixed state.

The main idea of \cite{Belin:2025wju} is to perform OPE contractions directly in the state, and use the known origin of the contractions in the full Euclidean geometry as a guiding principle. Starting again from our half genus-5 state
\be
\ket{\frac{1}{2} [g=5]}=\sum_{i,j,k,l,m,n,o} C_{ikl}C_{kmn}C_{lmo}^*C_{noi}^* \mathbb{O}_{i,j}(\Omega,h_k,h_l,h_m,h_n,h_o) \ket{0}_L \otimes  \ket{0}_R \,,
\ee
one could try and do OPE contractions (which put different triplets of indices equal to each other) between the four OPE coefficients of the state. This is essentially what was done in \cite{Belin:2025wju}. Here, we see that this is not the right logic to follow: the wormhole which leads to a baby universe does not come from OPE contractions within the ket state, but rather comes from OPE contractions between some OPE coefficients in the bra state and others in the ket. We must therefore try something different.

We will see that what works is to consider the density matrix of the state, and perform OPE contractions there. This might seem trivial, since the state at hand is pure, but just like in \cite{Belin:2025wju}, the OPE contractions transform the state. There, they transformed an unentangled state into an entangled one. Here, they will transform a pure state into a mixed one. Let us see this explicitly.

We start with the half genus-5 state in the low-temperature limit
\be
\ket{\frac{1}{2} [g=5]}\approx \sum_{l,m,n} |C_{lmn}|^2\mathbb{O}_{\mathbbm{1},\mathbbm{1}}(\Omega,h_l,h_m,h_n,) \ket{0}_L \otimes  \ket{0}_R \,.
\ee
We then write the density matrix.
\be \label{defrhopure}
\rho_{\text{pure}}=\ket{\frac{1}{2} [g=5]}\bra{\frac{1}{2} [g=5]} = O \ket{0}_L \otimes  \ket{0}_R \bra{0}_L \otimes  \bra{0}_R  O^\dagger \,,
\ee
with
\be
O=\sum_{l,m,n} |C_{lmn}|^2\mathbb{O}_{\mathbbm{1},\mathbbm{1}}(\Omega,h_l,h_m,h_n,) \,.
\ee
Recall that the operator $O$ is built from the identity operator along with all Virasoro modes, and the coefficient in front of these distinct operators care about the energy of the heavy states running in the loops of the internal  genus-2 surface.

The main trick now is to perform an OPE contraction on the density matrix $\rho_{\text{pure}}$. We will denote this operation by
\be
\rho_{\text{pure}}\longrightarrow\rho  \,.
\ee
Just like in \cite{Belin:2025wju}, we emphasize that this operation is not about ensemble averaging the theory or even coarse-graining the state. It is just a microscopic prescription on how to define the CFT state. It is of course guided by what we have learned about 3D gravity in terms of averaging/coarse-graining, but here it is a prescription that can be applied microscopically to the density matrix $\rho_{\text{pure}}$ in any given CFT. We also emphasize that while $\rho_{\text{pure}}$ was prepared with a CFT Euclidean path integral, this will no longer be the case of $\rho$. This will be important when we discuss fluctuations shortly.

The output of this procedure, using the Gaussian approximation for OPE coefficients gives
\be \label{rhodef}
\rho= \sum_{l,m,n} |C_{lmn}|^4 \mathbb{O}_{\mathbbm{1},\mathbbm{1}}(\Omega,h_l,h_m,h_n,) \ket{0}_L \otimes  \ket{0}_R \bra{0}_L \otimes  \bra{0}_R \mathbb{O}^\dagger_{\mathbbm{1},\mathbbm{1}}(\Omega,h_l,h_m,h_n,)  \,.
\ee
The difference between this state and $\rho_{\text{pure}}$ is that the heavy states running in the bra and ket parts of the density matrix are the same here, while they were distinct in $\rho_{\text{pure}}$. They have been set to be the same by the OPE contraction. One can now check that this state is no longer pure. Before, it could be written as in \rref{defrhopure}, but this is no longer the case, due to the OPE contraction.

One can easily check that the state produces the non-handlebody at genus-5\footnote{Again, this can be checked explicitly for pure gravity. We briefly comment on the addition of matter in the discussion.}
\be
\tr \rho = Z_{\text{non-handlebody}} \,,
\ee
by construction. As such, we can convincingly argue that $\rho$ is the CFT state dual to the geometry that includes a genus-2 baby universe.

It is also important to note that there is no AR puzzle for this state: the CFT state has as its dominant geometry a baby universe, so the state of the bulk matter on the two AdS spaces should be mixed. But indeed, we see explicitly that the CFT state is mixed.

\subsection{The baby universe is semi-classical}

We will now argue that in the construction described above, the baby universe is actually semi-classical. For this, we will study the fluctuations of the wave-function coefficients as was proposed in \cite{Kudler-Flam:2025cki}.\footnote{Note that our state is not properly normalized, so in principle one should worry about the fluctuations of the normalization. Those can also be shown to be exponentially small.}

In that context (see also \cite{Liu:2025cml}), the issue at hand was the convergence and/or fluctuations of the wave-function coefficients in the strict large $N$ limit. Here, we will not worry about the strict large $N$ limit, but simply analyze the size of the fluctuations for the wave-function coefficients directly, at large but finite $N$. Unlike the setup involving shell geometries, we will see that the fluctuations are small. This can be tracked back to the fact that the baby universe state $\rho$ is \textit{not} a state prepared by the CFT path integral.

Let us consider the quadruple vacuum wave-function coefficient
\be
A_{0,0,0,0}\equiv \bra{0}\otimes \bra{0} \rho \ket{0}\otimes \ket{0} \,.
\ee
This can be computed from \rref{rhodef}, and reads
\be
A_{0,0,0,0} \approx q_L^{-c/12} q_R^{-c/12} Z_{MM} \,.
\ee
The two factors $q_{L,R}^{-c/12}$ (recall $q\sim e^{-2\pi\beta}$) correspond to the four long half-tubes (left and right, bra and ket), where the vacuum is propagating. The rest of the wave-function coefficient is simply the partition function on the Maoz-Maldacena wormhole geometry\footnote{This partition function can be matched explicitly to the Maoz-Maldacena geometry in the case of pure gravity \cite{Belin:2020hea,Chandra:2022bqq}.}
\be
Z_{MM}=\sum_{l,m,n} |C_{lmn}|^4 |\mathcal{F}_{g=2}(\Omega)|^4 \,.
\ee
By looking only at the vacuum wave-function coefficient, we have essentially pinched the genus-5 surface into a product of two genus-2 suraces, which is why the genus-5 Virasoro blocks have been reduced to genus-2 blocks and why the Maoz-Maldacena partition function appears directly.

We would now like to study the square of this quantity, and see if the variance is large. We need to compute
\be
A_{0,0,0,0}^2 \Big|_{\text{connected}} \,.
\ee
While every calculation of the paper so far did not require explicit coarse-graining or averaging, this calculation can only be done from a coarse-grained perspective. In any given theory, $A_{0,0,0,0}$ is just a number and the connected square gives zero. So we can either view this number as varying as we vary $N$, which is the perspective taken in \cite{Kudler-Flam:2025cki,Liu:2025cml}, or study it from a coarse-grained perspective. We will choose the latter.

We have
\be
A_{0,0,0,0}^2 = q_L^{-c/6} q_R^{-c/6} \sum_{l,m,n} |C_{lmn}|^4 |\mathcal{F}_{g=2}(\Omega,h_l,h_m,h_n)|^4 \sum_{i,j,k} |C_{ijk}|^4 |\mathcal{F}_{g=2}(\Omega,h_i,h°j,h_k)|^4 \,.
\ee
The connected part will come from contracting OPE coefficients between the two factors of $A_{0,0,0,0}$. This will automatically reduce the number of sums from six to three, and as such, will be exponentially suppressed (approximately by $e^{-3S(E)}$ where $E$ is again fixed by the moduli $\Omega$):\footnote{This expression should be understood as schematic, the naive entropy counting due to indices being put equal to each other does not always lead to the naive suppression factor (see for example \cite{Belin:2021ryy,Anous:2021caj}), but in any case, there will be an exponential suppression in $c$.}
\be
A_{0,0,0,0}^2 \Big|_{\text{connected}} \sim e^{-3 S(E)} \,.
\ee
This means that the variance is small here, rather than order one as in the shell states \cite{Kudler-Flam:2025cki}.

The major difference comes from the fact that our state is not prepared by the CFT Euclidean path integral. If it were, then the combinatorial arguments of \cite{Kudler-Flam:2025cki} would be unavoidable. But our state is not determined by fixing boundary conditions for the gravity path integral. If we had fixed the boundary conditions to be two genus-2 boundaries connected by long tubes, then our geometry would contribute, but it would not be the dominant one. This is what we found in section \ref{sec:2}. But here, our state was made dominant by a particular microscopic prescription, and in the process, it is no longer prepared by a Euclidean path integral. This made the wormhole dominant, but also implies that the variance of the wave-function coefficients are small.

We have achieved what we set off to do: make the baby universe both dominant and semi-classical. In \cite{Liu:2025cml,Kudler-Flam:2025cki}, it was proposed to achieve this by averaging over $N$ (i.e. over $c$). This was not necessary for us, although it would be very interesting to see if one can interpret our microscopic prescription as being equivalent to averaging over $c$ in some way. The price to pay was that the CFT state became mixed, but this is expected both from the point of view of \cite{Kudler-Flam:2025cki}, and more intuitively from the bulk since the presence of the baby universe makes the bulk state mixed.

\subsection{Interpretation of the baby universe state and VTQFT}

An immediate question is the interpretation of the mixedness of the baby universe state. What is the state on the two AdS disks entangled with? In the discussion, we will discuss the case where bulk matter is added which would lead to a more straightforward answer: the Hilbert space that the two AdS disks are entangled with corresponds to the Hilbert space of bulk matter fields living in the baby universe. If the bulk theory is pure gravity, then the situation is more subtle as pure gravity is topological and there are no local degrees of freedom living in the closed universe.

For the case of pure gravity, the answer is provided by Virasoro TQFT (VTQFT) \cite{Collier:2023fwi}. When slicing the bulk as in Fig. \ref{fig:non-handlebody}, we can ask what the VTQFT amplitudes teach us about the bulk state. We see that AdS disks, which extend to AdS boundaries, give vectors in the Hilbert space of the vacuum Verma module. These can be traded for AdS half-ball boundaries with operators inserted at the northern pole, whether the identity or powers of the stress-tensor and its derivatives.\footnote{One could also ask a more general question, where non-trivial Wilson lines are allowed to flow through the vertical cylinders, but we will not discuss this possibility here.} The asymptotically AdS half genus-5 boundary should be viewed as summing half genus-5 OPE blocks, where the internal operators that run through the cycles of the genus-2 surface are being summed over. Finally, there is the finite boundary, which is the hashed genus-2 closed surface: the baby universe. Doing the VTQFT path integral with such a boundary produces a state in the Hilbert space of genus-2 conformal blocks. These states are labeled by a choice of three operators (three Virasoro representations). This should be distinguished from the case of asymptotically AdS boundaries where the boundary conditions give a sum over conformal blocks, with prescribed moduli. Let us call vectors in this Hilbert space
\be
\ket{i, j, k} \,.
\ee
This means that the VTQFT amplitude on the slicing of the non-handlebody produces a state
\be
\sum_{i,j,k,l,m,n} a_{i,j,k,l,m,n} |C_{lmn}|^2 \mathbb{O}_{\mathbbm{1},\mathbbm{1}}(\Omega,h_l,h_m,h_n,) \ket{0}_L \otimes  \ket{0}_R \otimes \ket{i,j,k} \,,
\ee
with particular wave-function coefficients $a_{i,j,k,l,m,n}$. VTQFT then predicts that the heavy operators running in the OPE block are the same as those on the finite size cut where the genus-2 baby universe lives. This means 
\be
 a_{i,j,k,l,m,n}=\delta_{i,l} \delta_{j,m} \delta_{k,n} \,.
\ee
The total states then becomes
\be
\sum_{l,m,n}  |C_{lmn}|^2 \mathbb{O}_{\mathbbm{1},\mathbbm{1}}(\Omega,h_l,h_m,h_n,) \ket{0}_L \otimes  \ket{0}_R \otimes \ket{l,m,n} \,.
\ee

It is worth commenting on the nature of the Hilbert space whose vectors we write as $\ket{i,j,k}$. This Hilbert space is \textit{not} the CFT Hilbert space. It is an auxiliary Hilbert space, just like the Hilbert space associated to any TQFT path integral with closed and finite boundaries. We see that here, it plays the role of an auxiliary system with which the two AdS universes can entangle themselves (see also \cite{Torres:2025jcb} for a related discussion). The situation is not unlike cutting open TQFTs to compute entanglement, where even if the Hilbert space is one dimensional on the global state, extra degrees of freedom can emerge at the entangling surface (see \cite{Shaghoulian:2023odo} for a nice discussion on this). It is very tempting to associate this auxiliary Hilbert space to that of an observer. The extent to which we can declare that the baby universe is \textit{encoded} in the CFT state remains unclear. The entanglement pattern between the two asymptotically AdS regions contains some knowledge of the auxiliary Hilbert space state, and we can prepare different states in that Hilbert space by changing the moduli of the Riemann surface. Topological operators in the baby universe would need to map to moduli changing operators from the point of view of the CFT. But we leave for future work a quantitative analysis of the encoding.

\begin{figure}[h]
\centering
  \includegraphics[width=.8\linewidth]{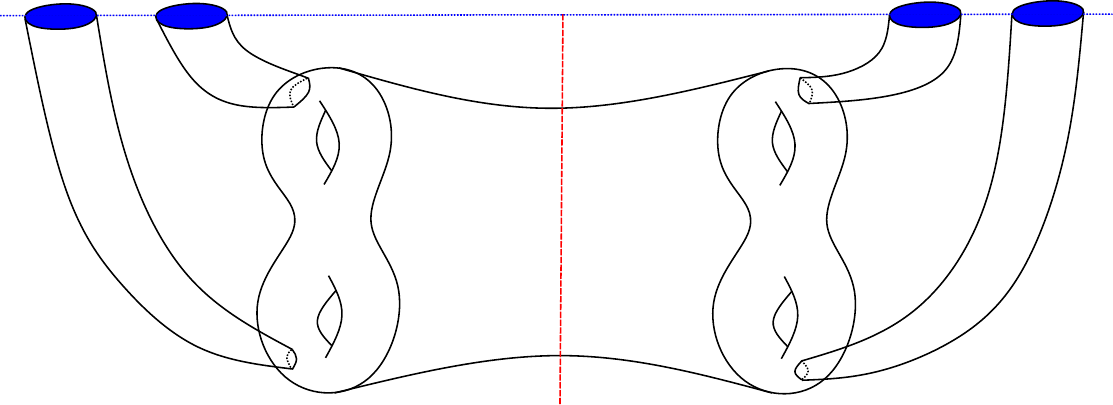}
\caption{A picture of the 3-geometry which prepares the canonical purification of the baby universe state. We have a Maoz-Maldacena wormhole, with each asymptotic genus-2 boundary having two cylinders attached to it. The topology is such that the four cylinders are filled, and the time-slice drawn as a blue cut has four disks. The left two disks are those of the original baby universe state, while the two other disks are those coming from the purification. The time-cut of the original baby-universe state, before being purified, corresponds to the union of the left blue segment until it reaches the red cut, and then it extends along the red cut. The genus-2 surface on the red cut is a sort of bottleneck for entanglement between the left and right sides.}
\label{fig:purification}
\end{figure}

We have seen that the state we obtain is an entangled state, where two AdS regions are entangled with an auxiliary system, which hosts its own auxiliary Hilbert space. Alternatively, we can trace over this auxiliary system and consider the density matrix which describes the two AdS spaces. It is interesting to ask what type of purifications exist. In the TQFT perspective, we purified the state through the auxiliary Hilbert space spanned by the heavy Virasoro representations. We will now discuss another purification, often referred to as the canonical purification (or GNS construction). This canonical purification  produces a state  $\ket{\psi_{\text{c.p.}}}$ in the double-copy Hilbert space, so it lives in
\be
\ket{\psi_{\text{c.p.}}} \in \mathcal{H}_{\text{CFT}}^{\otimes 4} \,.
\ee
In this case, the canonical purification yields the state
\bea
\ket{\psi_{\text{c.p.}}}&=& \sum_{l,m,n} |C_{lmn}|^4 \mathbb{O}_{\mathbbm{1},\mathbbm{1}}(\Omega,h_l,h_m,h_n,) \ket{0}_{L^1} \otimes  \ket{0}_{R^1} \notag \\
&\otimes & \Big( \mathbb{O}_{\mathbbm{1},\mathbbm{1}}(\Omega,h_l,h_m,h_n,)\Big|_{\text{CPT}}\ket{0}_{L^2} \otimes  \ket{0}_{R^2} \Big) \,.
\eea
It is also easy to draw a picture for the manifold on which to perform the VTQFT path integral to produce this state. We draw it in Fig. \ref{fig:purification}.

We see that the path integral is done on a Maoz-Maldacena wormhole with four solid cylinders sticking out. If we cut the picture in two, resulting in the description of the time-slice, we replace the two purifying asymptotic AdS regions by the closed genus-2 baby universe. So from this perspective, the baby universe is in some sense an entanglement bottleneck between the four asymptotic universes. It would be interesting to make this more precise.

\section{Discussion \label{sec:4}}

In this paper, we have presented a setup to study ideas related to baby universes, the size of their Hilbert space, and the puzzle raised by Antonini and Rath \cite{Antonini:2024mci}. The setup is to consider states prepared by the Euclidean path integral on higher genus surfaces in holographic 2D CFTs, and the role of topology is to replace the shell states of the AS${}^2$ construction. We found that baby universes exist, but in states prepared by the CFT path integral, the geometries where they exist are exponentially suppressed compared to geometries where they are absent. This implies that they do not offer a semi-classical description of the CFT states, since they only exist in an exponentially small piece of the wave-function. We also implemented a prescription introduced in \cite{Belin:2025wju}, to make the baby universe dominant. This can be accomplished for the states we studied, but the relevant states are no longer prepared by a CFT path integral. The resulting state is mixed, as expected from the bulk due to the presence of the baby universe, and there is thus no AR puzzle. We also checked that the fluctuations in the wave function coefficients were small, confirming that in this setup, the baby universe is semi-classical. We conclude with some open questions.

\subsection*{Adding matter}

Many of the steps in this paper were quite general, and can be applied to any holographic CFT. For example, the baby universe density matrix \eqref{rhodef} can be considered in any holographic CFT - like the D1D5 CFT at strong coupling - and we believe it to accurately be described by the baby universe geometry. That being said, we have only been able to check this explicitly in the case of pure gravity. This is because in pure gravity, the partition function on the Maoz-Maldacena wormhole can be computed explicitly and exactly in $c$. In a holographic dual with matter, this has not been achieved yet. It would require correcting the statistical distribution of OPE coefficients, and keeping contributions beyond the crossing transformations of the identity operator. Many more features can appear once light matter is included. For example, even the handlebody geometries are not necessarily stable if matter is sufficiently light \cite{Belin:2017nze,Dong:2018esp}.

The addition of bulk matter would actually make certain features of our setup more trackable. For example, the Hilbert space associated to the closed universe is confusing in pure gravity, since it does not really correspond to any local degrees of freedom of the closed universe - there are none in a topological theory. With matter, the Hilbert space is simply the Hilbert space of a local QFT on the closed universe and hence it would be interesting to see how this is reflected in the CFT state. In \cite{Liu:2025cml}, a proposal is put forward on how to think about the algebra of operators in the pure Maoz-Maldacena cosmology, in terms of the quotient used to obtain the wormhole from $\mathbb{H}^3$. It would also be interesting to see if one can connect those ideas to our setup.

\subsection*{More complicated baby universes}

In this paper, we discussed a simple example stemming from a genus-5 non-handlebody whose Lorentzian slicing gives a genus-2 closed baby universe. We picked this example for concreteness, But it can easily be generalized to include either higher genus baby universes, or even multiple baby universes of various genera. The construction will always involve a CFT state prepared on some higher topology, and the different topologies that we can obtain depend on the type of contraction between the bra and the ket. If there are four handles in the state preparation, we can imagine having either a closed universe of genus four, or alternatively, two closed universes of genus two. We see no obstruction to finding mixed states described at leading order by any of these geometries, but it would be interesting to establish this more firmly.

An interesting question is the fate of our construction at very large genus. As the genus becomes sufficiently large (scaling with $c$), a semi-classical description of the bulk will most likely disappear (see \cite{Belin:2023efa} for a discussion on this). But if we ignore this issue, it is possible that the physics drastically changes. The handlebody geometries will still be leading compared to any single connected (i.e. non-handlebody) geometry, but the number of connected geometries could eventually start to overcome the suppression. It is possible that in the right limit, we can get connected geometries as the leading order contribution. If true, this would present an interesting analogy to shell states, where the genus of the surface may serve as a proxy for the number of operator insertions that make the shell. It would be interesting to study this in more detail.

\subsection*{Other saddles}

Another interesting question to pursue is to perform a more in-depth study of the sum over topologies contributing to the overlap
\be
\braket{\frac{1}{2} [g=5] | \frac{1}{2} [g=5]}  \,.
\ee
There is a whole infinite set of solutions to Einstein's equations with genus-5 boundary, and a systematic description of these solutions in relation to the statistics of OPE coefficients will be presented in \cite{Belin:2026pko}. One interesting question would be to understand how to think about other non-handlebody manifolds, and what their Lorentzian slicings look like. It appears clear from our construction that a necessary condition for the appearance of a baby universe is that it contributes to a connected two-point function of genus-2 partition functions, ala Maoz-Maldacena. Some non-handlebodies will simply correspond to more complicated topologies contributing to a single genus-2 partition function. But does an arbitrary connected geometry, embedded in the genus-5 boundary case, lead to a baby universe? Can the baby universe be made dominant by a prescription similar to that described in section \ref{sec:3}? Does it always correspond to a Virasoro TQFT amplitude with a finite boundary higher genus surface (or a union of higher genus surfaces?)

The reverse question is also interesting: given two AdS disks touching a half genus-5 surface at an AdS boundary, along with a set of finite closed boundaries. Can a Virasoro TQFT amplitude always be found by connecting the two using a suitable surgery operation? Does it correspond to the $\mathbb{Z}_2$ slicing of a smooth Euclidean geometry? It would be interesting to explore these questions further.

\subsection*{Over-completeness and the finiteness of the black hole Hilbert space}

Another interesting idea to pursue would be to study another situation where thin shells have been successfully used: the over-completeness of the gravitational states and the finiteness of the black hole entropy \cite{Balasubramanian:2022gmo}. This is similar to the setup here, but in the regime before the Hawking-Page transition described by equation \eqref{highergBH}. The overcomplete family of CFT states could be taken either as Einstein-Rosen bridges with higher and higher topology behind the horizon, or one could restrict to genus-2 but consider a multitude of different points on the moduli space of the genus-2 surface.\footnote{We would like to emphasize that it is not clear whether one can really reliably pick $e^S$ different semi-classical states in this way and still have a trust-worthy semi-classical treatment of all overlaps.}

This setup would provide an interesting explicit alternative framework to study over-completeness. Shell geometries are such that the overlap between two different shell geometries $S_1$ and $S_2$ is zero. For many different shells $S_j$, the overlaps satisfy
\be
\overline{\braket{S_i | S_j}} = \delta_{i,j} \,.
\ee
This is not the behaviour expected of semi-clasiscal states: semi-classical states should have non-zero (but exponentially small) overlaps with each other, just like coherent states have non-trivial overlap with each other in quantum mechanics. This is even true for states that correspond to different topologies on the spatial slice \cite{Jafferis:2017tiu}. In general, for two different semi-classical states $\ket{\psi_{1,2}}$, we expect
\be
\overline{\braket{\psi_1 | \psi_2 }} = e^{-c f_{12}} \,,
\ee
for some order one function $f$ that depends on the details of the states.

It then becomes a complicated question to understand how these non-perturbatively small (but non-fluctuating effects) compete with the physics of wormholes. Our setup provides a clear but technically challenging path forward: one can study all the overlaps, whether the non-fluctuating piece or the higher-points of the overlaps corresponding to wormholes, in 3D gravity and there are explicit expressions to evaluate all of these quantities. It would be interesting to see whether one recovers the black hole entropy in this more explicit setup.

\subsection*{Closed universes without AdS universes}

An interesting limit to consider is to take the cylinders to be infinitely long. In that case, the states on the left and right AdS universes are frozen to be in the vacuum. Note that from the genus-5 partition function point of view, this is a pinching limit. What is the fate of the baby universe in such a limit? Because the AdS universes have been frozen to be in the vacuum, we are essentially in the setups discussed in \cite{Antonini:2022blk,Antonini:2022ptt,Betzios:2024oli} where there is just a closed universe and nothing else. How is the physics of this closed universe encoded?

Naively, the auxiliary Hilbert space labeled by three Virasoro representations still survives, and can be associated to this closed universe, but it is unclear exactly how to interpret it. Things become even more confusing if we try to add matter fields but insist on keeping a VTQFT perspective, albeit with new Wilson lines. The matter fields correspond to Wilson lines that can propagate in the bulk, which end on the boundary where the associated local operators are inserted. But if we insist on inserting such a Wilson line and have it go through the baby universe, then on the time-slice we are now getting one-punctured genus-2 surfaces. This corresponds to a \textit{different} (and larger!) auxiliary Hilbert space. So it appears as if probing the baby universe states requires constantly changing its Hilbert space, creating a large number of superselection sectors, which is quite confusing. Similar observations have been made once observers are added \cite{Shaghoulian:2023odo,Abdalla:2025gzn,Harlow:2025pvj}. It would be interesting to understand this better.

\section*{Acknowledgments}
It is a pleasure to thank Tarek Anous, Stefano Antonini, Lorenz Eberhardt, Tom Hartman, Kyriakos Papadodimas, Boris Post, Pratik Rath, Martin Sasieta, Edgar Shaghoulian, Julian Sonner, Alejandro Vilar Lopez and Tom Yildirim for interesting discussions. JdB is supported by the European Research Council under the European Unions Seventh Framework Programme (FP7/2007-2013), ERC Grant agreement ADG 834878. AB thanks the Centro de Ciencias de Benasque Pedro Pascal and the program "Gravity - new quantum and string perspectives" for hospitality, and for providing a vibrant atmosphere that enabled the beginning of this work.

\bibliographystyle{ytphys}
\bibliography{ref}

\end{document}